\documentclass[aps,twocolumn,groupedaddress,showpacs]{revtex4}
\usepackage{graphicx}
\usepackage{dcolumn}
\usepackage{bm}
\usepackage{epstopdf}
\usepackage[dvipsnames,usenames]{color}
\usepackage{color}
\usepackage{float}
\usepackage{amsmath}
\usepackage{amssymb}
\usepackage{amsthm}
\usepackage{mathrsfs}

\usepackage[T1]{fontenc}
\usepackage[skip=1ex]{caption}
\usepackage{booktabs,tabularx}
\usepackage{siunitx}

\renewcommand{\thefootnote}{*}
\emergencystretch=\maxdimen
\begin{document}
\title{Quasi-BIC laser enabled by high-contrast grating resonator for gas detection}
\author{Haoran~Zhang$^{1,2}$, Tao~Wang$^{1,2}$,  Jingyi~Tian$^{3}$, Jiacheng~Sun$^{1,4}$, Shaoxian~Li$^{1,2}$, Israel De Leon$^{5}$, Remo Proietti Zaccaria$^{6}$, Liang~Peng$^{1,2}$, Fei~Gao$^{7}$, Xiao~Lin$^{7}$, Hongsheng~Chen$^{7}$, Gaofeng~Wang$^{1,2}$}

\affiliation{$^1$Engineering Research Center of Smart Microsensors and Microsystems of MOE, Hangzhou Dianzi University, Hangzhou, 310018, China}
\affiliation{$^2$School of Electronics and Information, Hangzhou Dianzi University, Hangzhou, 310018, China}
\affiliation{$^3$Centre for Disruptive Photonic Technologies, TPI, SPMS, Nanyang Technological University, Nanyang, 637371, Singapore}
\affiliation{$^4$School of Zhuoyue Honors, Hangzhou Dianzi University, Hangzhou, 310018, China}
\affiliation{$^5$School of Engineering and Sciences, Tecnol\'ogico de Monterrey, Monterrey, Nuevo Le\'on 64849, Mexico}
\affiliation{$^6$Cixi Institute of Biomedical Engineering, Ningbo Institute of Materials Technology and Engineering, Chinese Academy of Sciences; Italian Institute of Technology via Morego 30, 16163 Genova, Italy}
\affiliation{$^7$Interdisciplinary Center for Quantum Information, College of Information Science and Electronic Engineering, Zhejiang University, Hangzhou 310027, China}

\date{\today}

\begin{abstract}
In this work, we propose and numerically investigate a two-dimensional microlaser based on the concept of bound states in the continuum (BIC). The device consists of a thin gain layer (Rhodamine 6G dye-doped silica) sandwiched between two high-contrast-grating layers. The structure supports various BIC modes upon a proper choice of topological parameters; in particular it supports a high-Q quasi-BIC mode when partially breaking a bound state in the continuum at $\Gamma$ point. The optically-pumped gain medium provides sufficient optical gain to compensate the quasi-BIC mode losses, enabling lasing with ultra-low pump threshold (fluence of 17 $\mu$J/cm$^2$) and very narrow optical linewidth in the visible range. This innovative device displays distinguished sensing performance for gas detection, and the emission wavelength sensitively shifts to the longer wavelength with the changing of environment refractive index (in order of $5 \times 10^{-4}$). The achieved bulk sensitivity is 221 nm/RIU with a high signal to noise ratio, and a record-high figure of merit reaches to 4420 RIU$^{-1}$. This ultracompact and low threshold quasi-BIC laser facilitated by the ultra-narrow resonance can serve as formidable candidate for on-chip gas sensor.
\end{abstract}


\maketitle 

\section{Introduction}
In recent years, the investigation of bound states in the continuum (BICs) has attracted substantial attention due to the interesting physics and practical applications~\cite{Joseph2021}. The BIC theory originated from quantum mechanics and was firstly proposed by von Neumann and Wigner in 1929~\cite{Neumann1929, Han2019}. Since then, it has been used to explain the important physical concept of resonance with infinite lifetime in various physical systems, such as photonics~\cite{Marinica2008, Hsu2013, Dong2021, Tang2021}, acoustics~\cite{Lyapina2015, Xiao2017} and water waves~\cite{Linton2007}. According to Neumann and Wigner's seminal work~\cite{Neumann1929}, when two resonances pass each other as a function of a continuous parameter, the two channels will interfere, and give rise to an avoided crossing for their resonances. Theoretically, at a given value of the continuous parameter, one of the channels vanishes entirely and hence becomes a dark mode (BIC mode) with an infinite quality ($Q$) factor~\cite{Azzam2018}. In practice, BICs are limited by finite structure size, material absorption, and structural imperfection~\cite{Koshelev2020}, they manifest themselves and collapse to Fano resonant states with long lifetime, also known as quasi-BICs~\cite{Hsu2016, Sadrieva2017}. Recently, high-$Q$ quasi-BICs have been observed in many passive systems, and it has been recognized that practical systems supporting such resonances are well suited for lasing and sensing applications~\cite{Kodigala2017, Ha2018}.
 
Sub-wavelength high-contrast gratings (HCGs) possess distinct features, such as broadband high reflectivity ($>99\%$) and high-$Q$ resonances ($>10^7$)~\cite{Karagodsky2012}. In particular, it has been shown that HCG systems can support BICs with improved spectral performance~\cite{Joseph2021, Lee2020}. The narrow spectral linewidth ($\gamma$) featured by these structures are very sensitive to changes in refractive index around them~\cite{Maksimov2020}, making them promising candidates for engineering optical sensors with high sensitivity ($S$) and excellent figure of merit, FOM = $S/\gamma$~\cite{Romano2018}. However, investigations thus far have considered only passive structures, which can limit significantly the device's sensitivity~\cite{Bosio2019}. Recent studies indicate that active sensors based on small lasers not only supply coherent radiation, but also show enhanced sensing performance~\cite{Ma2014, Sun2021}. Thus, photonic structures supporting high-$Q$ quasi-BIC lasers are of great interest in sensing applications, as they could offer possibilities of achieving a sensing performance beyond that of more conventional photonic sensors~\cite{Yesilkoy2019, Liu2017}.

Lasing action based on BICs or quasi-BICs in two-dimensional photonic-crystal structures has been reported recently~\cite{Kodigala2017, Ha2018, Huang2020}. Nonetheless, investigations on potential applications of such devices are scarce, e. g. active sensing based on BIC lasers. In this work, we propose an innovative resonator composed of highly reflecting HCG layers surrounding a thin organic-dye-doped SiO$_2$ layer working as gain material for laser emission. This ultracompact resonator design not only opens an alternative route to generate different resonance modes, but also provides a platform to demonstrate a Fabry-P\'erot quasi-BIC laser by combining organic dye molecules with inorganic grating. Finite difference-time-domain (FDTD) numerical simulations have been performed to evaluate the performance of the aforementioned system with the results suggesting that a high-$Q$ quasi-BIC mode ($Q > 10^4$) can be established by carefully optimizing the HCG structural parameters. Then, when the quasi-BIC mode spectrum overlaps with that of the gain material, a pronounced lasing action is obtained. More significantly, this quasi-BIC laser allows for a highly precise environment change detection through the reading of the emission wavelength shift. The achieved narrow linewidth (~0.05 nm) combined with a large bulk sensitivity of 221 nm/RIU, results in a FOM of 4420 RIU$^{-1}$, which is approximately 20 times larger than typical passive sensors~\cite{Maleki2020}. Therefore, our results suggest that the proposed quasi-BIC laser offers great potential for optical sensing of gases in low-concentration.  

\section{Resonator design and numerical model}
\subsection{Structure design}
\begin{figure}
\centering
\includegraphics[width=1.0\linewidth,clip=true]{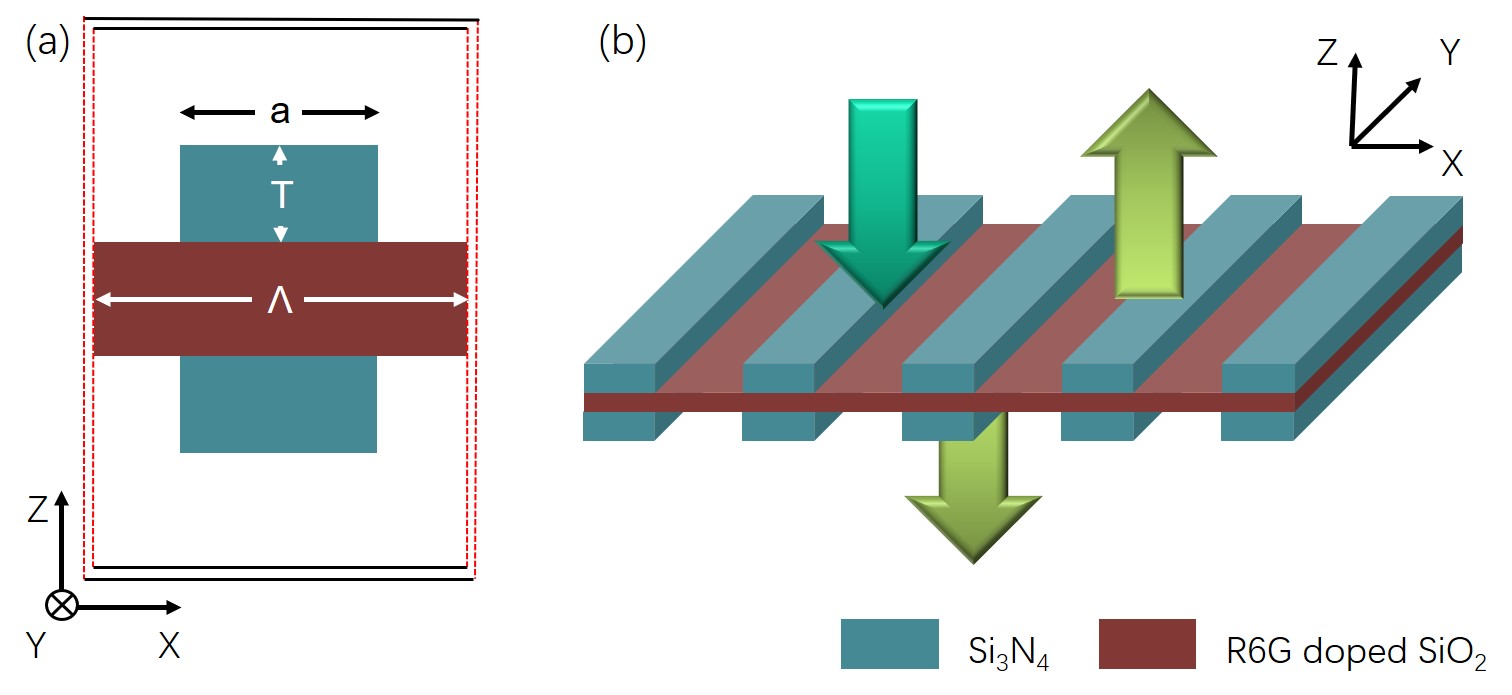}
\caption{Schematic of proposed resonator based on HCGs: (a) single unit cell of two dimensional resonator; (b) whole structure with the organic R6G thin layer. The dark green downward arrow represents optical pump, the light green arrows on both sides of resonator are emitted light from the resonator.}
\label{structure}
\end{figure}

The schematic diagram of proposed double-HCG resonant structure is shown in Fig.~\ref{structure}. The structure is designed to support a quasi-BIC at wavelength of 570 nm. The active region consists of 215 nm thick SiO$_2$ layer doped with Rhodamine 6G (R6G), as shown in Fig.~\ref{structure}(a). A similar gain medium has been reported in a previous work~\cite{Anedda2005}. The R6G layer is surrounded by Si$_3$N$_4$ stripes with a symmetric arrangement along $z$ direction, and the refractive index of Si$_3$N$_4$ is 2.0. The HCG parameters are described by the thickness of grating layer $T$, filling factor (the grating width divided by the period) $F$ and the period $\Lambda$. We consider the dimensions of whole device in $x$- and $y$-direction are infinite, as exhibited in Fig.~\ref{structure}(b). The detail parameters of materials are shown in Table 1. 

\begin{figure}
\centering
\includegraphics[width=1\linewidth,clip=true]{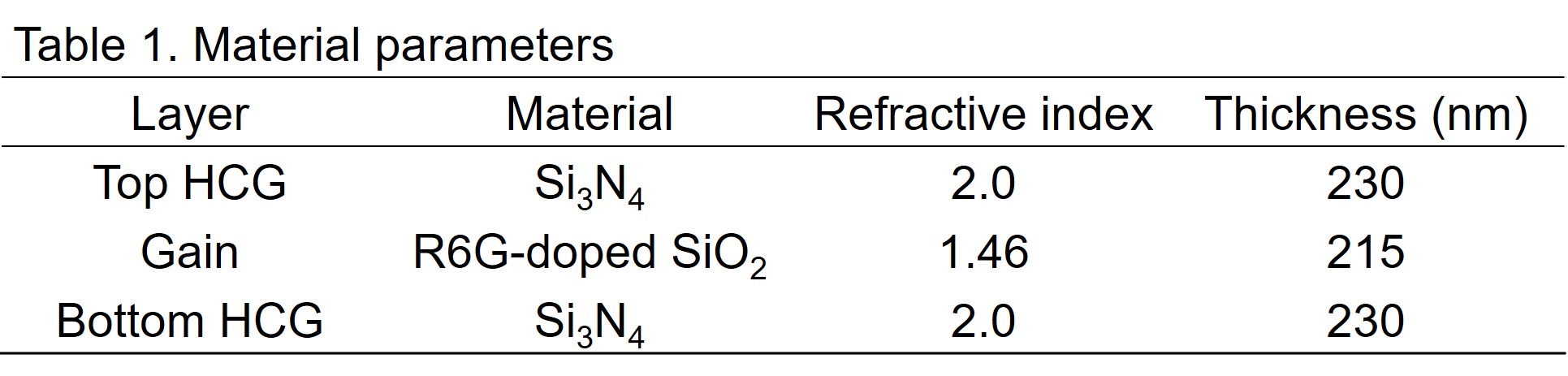}
\label{Ptable}
\end{figure}

A freestanding cascade grating fabrication technique, based on positive resist e-beam lithography (EBL) and ion coupling plasma, is supposed to be feasible to fabricate the designed structure. The SiO$_2$ layer doped with R6G can be made by using sol-gel technique, which is a suitable method for incorporating organic molecules into inorganic solid hosts. The low-temperature process involving the hydrolysis and condensation reactions of metal alkoxides enables us to dope different molecules with a poor thermal stability into coating films~\cite{Yanagi1998}. The fabrication process mainly includes thin film deposition and grating pattern. The Si$_3$N$_4$ thin film is uniformly coated with SiO$_2$ doped R6G sol solution, then cover another Si$_3$N$_4$ thin film on the other side. The grating structures can be separately patterned on both side by using EBL and ICP with CHF$_3$ gas~\cite{Hong2019}. A similar sandwiching structure consisting of two aluminum gratings and a Si$_3$N$_4$ membrane has been reported recently~\cite{Liang2018}.

\subsection{Theory and simulation}
A semi-quantum framework is adopted to simulate the interaction between the electromagnetic fields and gain medium. The organic dye molecules in this work are well described as a four-energy-level system, and the population inversion is generated between the electronic level of $L_2$ state and $L_1$ state. As shown in Fig.~\ref{fourlevel}, which presents the population dynamics of the four relevant electronic levels of the molecule, such as the ground state, $L_0$, and the three excited states, $L_1$, $L_2$, and $L_3$. The molecular polarization given by spontaneous and stimulated transitions are described through the following equation~\cite{Zhong2013}:

\begin{figure}
\centering
\includegraphics[width=0.90\linewidth,clip=true]{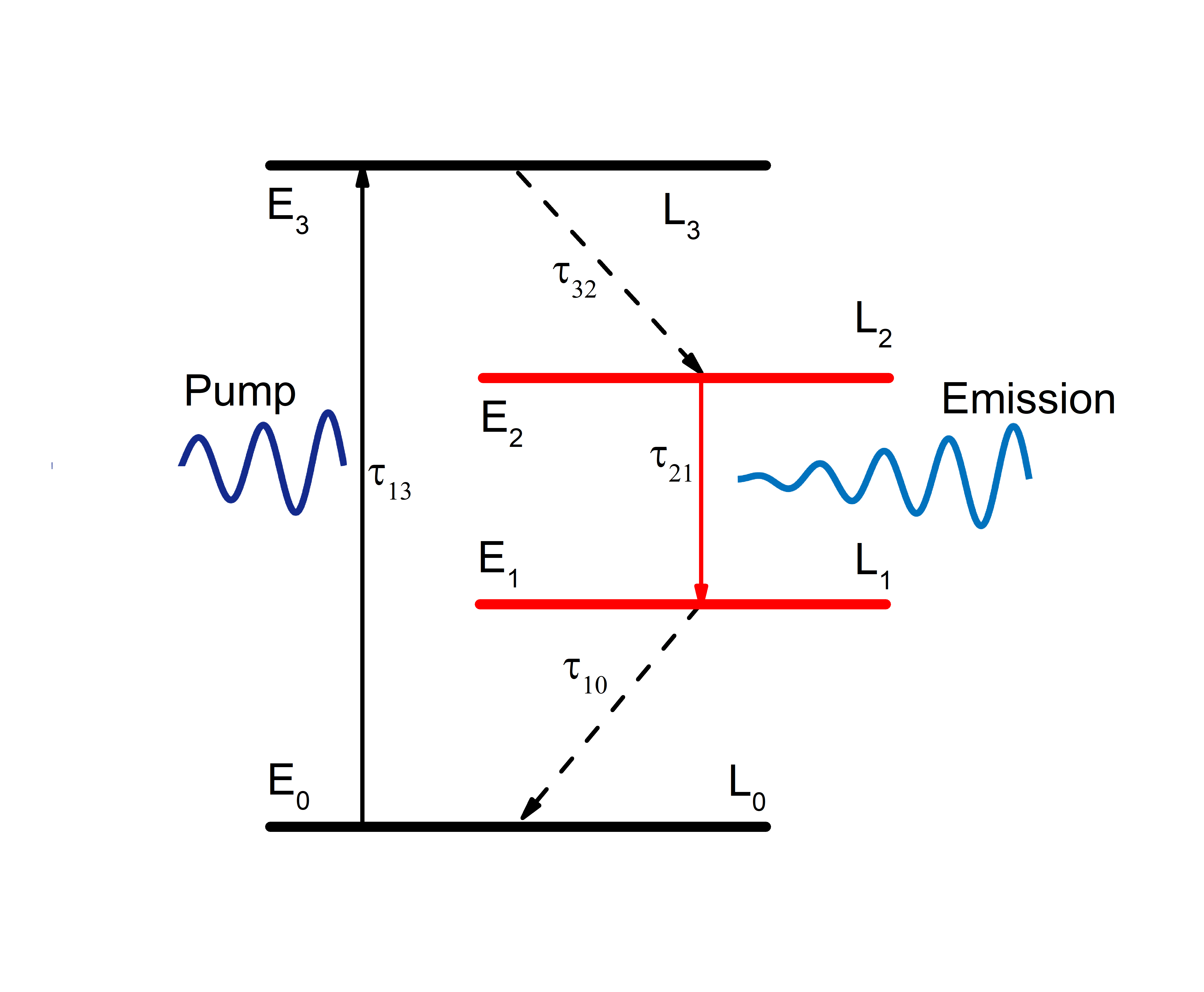}
\caption{Energy level diagram and parameters used for modelling the dye molecules in FDTD simulations. The transition energies are $E_{30} = E_3 - E_0$ and $E_{21} = E_2 - E_1$. Since the transition from $L_3$ to $L_2$ is much faster than the transition from $L_3$ to $L_0$, the excitation quickly decay to $L_2$, leading to a population inversion for the transition from $L_2$ to $L_1$.}
\label{fourlevel}
\end{figure}

\begin{eqnarray}
\frac{d^2\overrightarrow{P}_{i,j}}{dt^2} + \Delta\omega_{i,j} \frac{d\overrightarrow{P}_{i,j}}{dt} + \omega^2_{i,j}\overrightarrow{P}_{i,j} = \kappa_{(i,j)}\Delta N_{i,j}(t)\overrightarrow{E}(t)
\end{eqnarray} 

\noindent where $\Delta\omega_{i,j}$ and $\omega_{i,j}$ are the bandwidth and frequency of the transition between states $i$ and $j$, $\kappa_{i,j} = 6\pi\varepsilon_0c^3/(\omega^2_{i,j}\cdot\tau_{i,j})$~\cite{Wang2020a}, $\tau_{i,j}$ represents the lifetime of the spontaneous emission. $\Delta N_{i,j}(t)$ is the difference of population density between two energy states of the interest. $\overrightarrow{E}(t)$ is the total electric field, and can be calculated by solving the curl Maxwell equations~\cite{Dridi2013}:

\begin{eqnarray}
\nabla \times \overrightarrow{E}(t) = -\mu_0 \frac{\partial \overrightarrow{H}(t)}{dt}\, ,\\
\nabla \times \overrightarrow{H}(t) = \varepsilon \frac{\partial \overrightarrow{E}(t)}{\partial t} + \frac{\partial (\overrightarrow{P}_{30}(t)+\overrightarrow{P}_{21}(t))}{\partial t}
\end{eqnarray}

In order to solve for the fields in a self-consistent fashion, it is necessary to couple the molecular polarization $\overrightarrow{P}_{i,j}$ to the electromagnetic fields via the rate equations of the four-level system describing the gain medium~\cite{Trivedi2017}. Such rate equations describe the time evolution of the energy states' population densities, and are given by Ref.~\cite{Zhou2013}. The time evolution of each energy state density is described by their respective spontaneous decay processes $N_i/(\tau_{i,j})$ and stimulated processes ($\overrightarrow{E}(t)\cdot d\overrightarrow{P}_{i,j}/dt$). The field involved in the rate equations is the total field and accounts for the effects of any local optical intensity on the dynamics of the population densities. 



The FDTD method is adopted to solve the coupled equations. For the case of R6G-doped SiO$_2$, we set that the absorption transition: $\lambda_a$ = 512 nm and $\Delta \lambda_a = 35$ nm; the emission transition: $\lambda_e = 570$ nm and $\Delta \lambda_e = 50$ nm; the concentration of dye molecule: $C = 1 \times 10^{19}$ cm$^{-3}$, which is comparable to the concentrations previously reported in the same material host~\cite{Anedda2005}; and the lifetimes: $\tau_{32} = \tau_{10} = 5 \times 10^{-14} $s, $\tau_{30} = 1 \times 10^{-9}$ s, $\tau_{21} = 1.8 \times 10^{-9}$ s. Finally, the calculated emission cross section is around 3 $\times$ 10$^{-16}$ cm$^2$, which is also consistent with the experimental report~\cite{Grossmann2010}.

\section{Bound states in the continuum and quasi-BIC lasing}
\subsection{BIC modes supported by the passive structure}
\begin{figure}
\centering
\includegraphics[width=1.0\linewidth,clip=true]{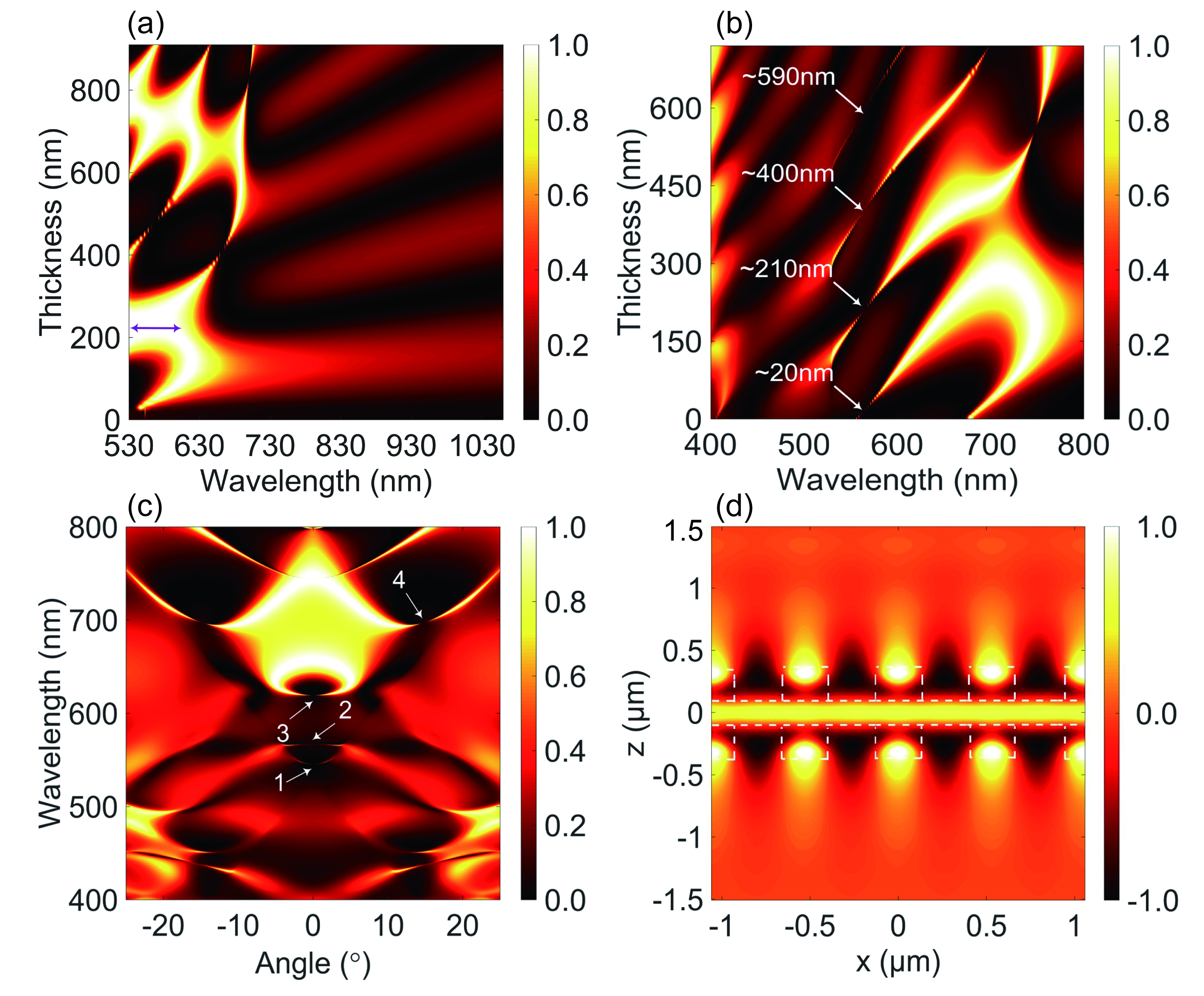}
\caption{(a) Reflectance spectra mapping as a function of grating thickness when the filling factor $F$ = 0.50; (b) reflectance spectra as a function of the thickness of SiO$_2$; (c) reflectance contour distribution of the architecture by sweeping the incident angle from -25$^{\circ}$ to 25$^{\circ}$; (d) electric field distribution profiles within four periods corresponding to the places marked by `2'.}
\label{Passive-typical}
\end{figure}

First, the optical properties of single HCG layer are analyzed. To obtain a perfect mirror for light trapped in a Fabry-p\'erot BIC system, we carry out FDTD simulations and calculate the structure's reflectance, for a normally incident plane wave polarized along the y-direction. Fig.~\ref{Passive-typical}(a) illustrates the results of our simulations as a function of the wavelength and of the grating thickness, taking a grating filling fraction $F$ = 0.5. We observe two regions, which we refer to as the dual/multi-mode region (from $\lambda$ = 530 nm to 700 nm) and the single-mode region ($\lambda > 700$ nm), as suggested by previous nomenclature~\cite{Hasnain2012}. For the single mode region, the grating operates in a longer wavelength regime, behaving like a quasi-uniform layer~\cite{Karagodsky2012}, so only Fabry-P\'erot modes are supported in this wavelength range. On the other hand, in the dual/multi-modes region, the pattern is quite different, describing contrasting regions with high ($> 99\%$) and low reflectance. In addition, we find the localized patterns don't change in an obvious manner when modify the filling factor $F$, instead, just an overall shifting to the longer wavelength. The reflectance spectrum shown in Fig.~\ref{Passive-typical}(a) reveals that with a proper selection of the grating parameters (e.g. the thickness $T$, filling factor $F$ and period $\Lambda$), an ultra-high ($> 99\%$) reflectivity over a spectral range is achievable (as shown by the double arrow line). 

We proceed now to study the optical properties of the entire passive structure (SiO$_2$ layer sandwiched by the two HCGs) as a function of the main structural parameters. Fig.~\ref{Passive-typical}(b) displays the reflectance spectra mapping as function of SiO$_2$ layer thickness, under the condition of $T$ = 230 nm, $F$ = 0.50, and $\Lambda$ = 530 nm. We notice there are a series of discrete reflectance vanishing points at the SiO$_2$ layer thickness of 20, 210, 400, and 590 nm, corresponding to a periodical disappearance of resonance. These singular points break the high reflection continuum, showing the typical feature of Fabry-P\'erot BIC. As mentioned before, a HCG layer possesses high reflectivity within a wide range, thus two HCGs layers with a symmetric alignment can make a Fabry-P\'erot cavity. Therefor, BICs are formed when the distance between two HCG layers is tuned to make the round-trip phase shifts add up to an integer multiple of 2$\pi$~\cite{Hsu2016}.

To obtain a quasi-BIC mode at wavelength of 566 nm that could be employed in a practical device, we slightly increase the thickness of SiO$_2$ layer by 5 nm. Fig.~\ref{Passive-typical}(c) shows the band diagram of the designed structure, which is plotted by varying the angle of incidence ($\theta$, from -25$^\circ$ to 25$^\circ$) with respect to the $z$-axis. We observe symmetric patterns relative to the axis of $\theta$ = 0$^\circ$ (also $\Gamma$ point), and the quasi-BIC mode at the position of mode 2, where is marked by the arrow and number `2', is successfully identified. This quasi-BIC mode exhibits a low dispersion near $\theta = 0^\circ$ around 566 nm, suggesting a strong resonance in individual unit cell~\cite{Taghizadeh2017}. Apart from mode 2, we also find other BIC modes at $\theta$ = 0$^\circ$ (e. g. positions marked by `1'and `3') and 14.8$^\circ$ (e. g. position marked by `4'). Those BICs feature vanishing linewidths, suggesting a non-leaky state~\cite{Marinica2008}. 

To further characterize the quasi-BIC at mode 2, we plot the field distribution profile in ($x$, $z$)-plane for this mode under the incident angle $\theta$ = 0$^\circ$, as shown in Fig.~\ref{Passive-typical}(d). The $E_y$ distribution of mode 2 over the ($x$,$z$)-plane shows a strong confinement within the cavity and a symmetric profile along $z$-direction. Due to the high reflectance of the HCGs and the sub-wavelength thickness of the SiO$_2$ layer, the first-order standing wave (i.e. one peak across the SiO$_2$ layer) is observed~\cite{Fitzgerald2021}. Typically, Fabry-P\'erot BIC can be generated in systems with two identical resonances coupled to a single radiation channel~\cite{Hsu2016}. Here, the round-trip phase shift between the two resonances in HCGs is 2$\pi$, thus the field is constructively interfered and trapped in the cavity while the resonant radiations from the HCGs interfere destructively out side of the cavity, which leads to the formation of Fabry-P\'erot BIC.

\subsection{Quasi-BIC lasing and characterization}
\begin{figure}
\centering
\includegraphics[width=1.0\linewidth,clip=true]{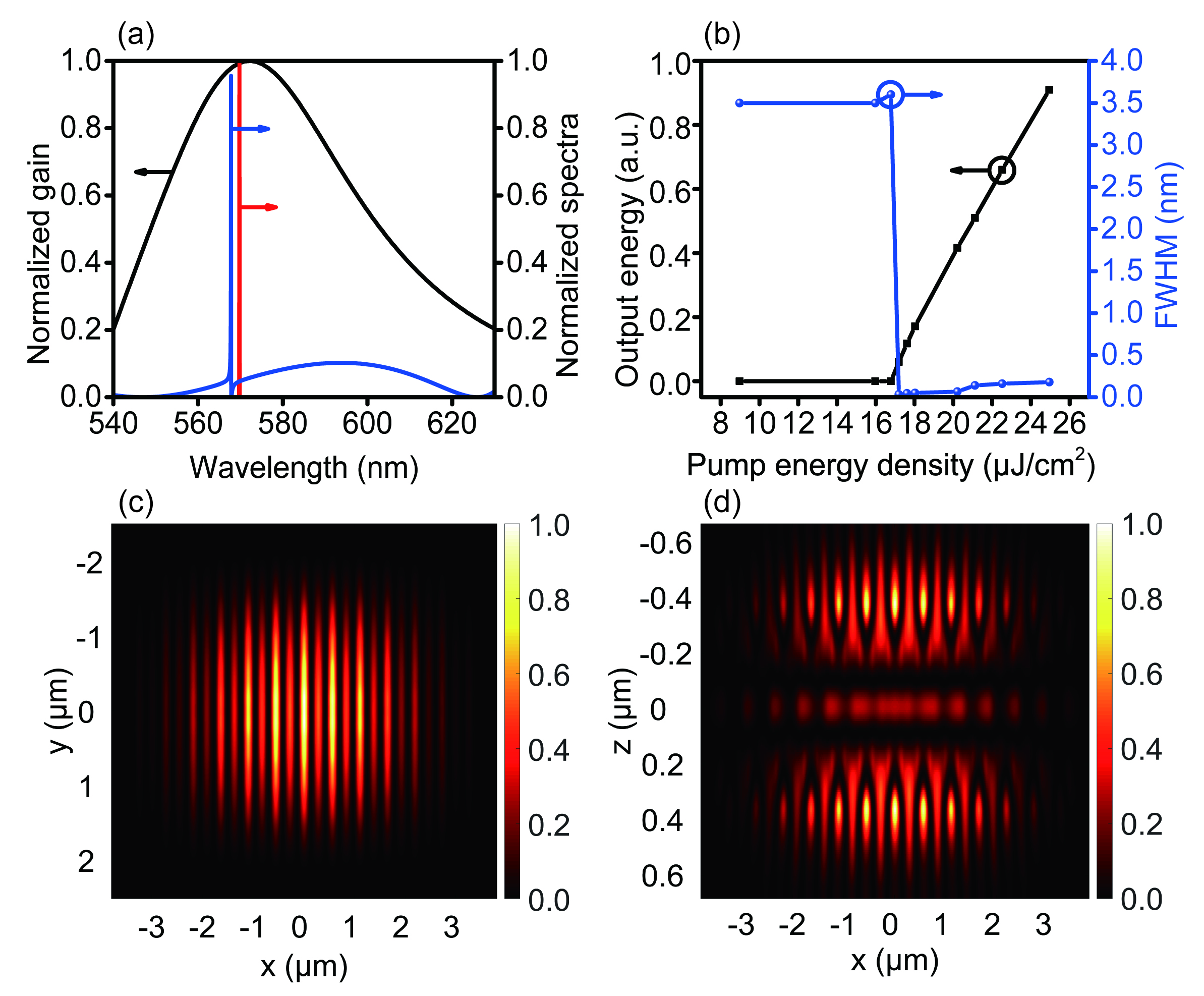}
\caption{(a) Gain spectrum of R6G (black curve), cavity mode spectrum (blue curve), and typical emission spectrum of laser (red curve) operated above threshold; (b) input-output function curve (black) and spectrum linewidth changing as a function of input energy (blue); (c) electric field intensity distribution of laser emission in ($x$, $y$) plane; (d) electric field intensity distribution in ($x$, $z$) plane.}
\label{Active}
\end{figure}

To realize quasi-BIC lasing, we selected the grating parameters ($T$ = 230 nm, $F$ = 0.50, and $\Lambda$ = 530 nm) which can provide the highest possible reflectivity (as marked in Fig.~\ref{Passive-typical}), and the SiO$_2$ thickness is 215 nm. Then, R6G dye molecules are doped into SiO$_2$ thin layer to compensated the internal losses of resonator. As shown in Fig.~\ref{structure}(b), the proposed sandwiched structure recalls conventional VCSELs, however our design significantly reduces the cavity length. Fig.~\ref{Active}(a) shows the gain spectrum of R6G (black curve), the spectrum of the cavity mode (blue curve) and typical lasing spectrum of resonator (red curve). Clearly, the R6G gain spectrum has a broad linewidth with its central region overlapping with the cavity mode peak, a feature guaranteeing a proper compensation of the cavity losses. Interestingly, the cavity mode shows the typical asymmetric line shape of a Fano resonance. This resonance is caused by the interference of radiative (bright) and non-radiative (dark) modes~\cite{Limonov2017, Heo2019}. For Si$_3$N$_4$-based HCG resonator, dark mode is associated with the Bragg scattering induced by HCG in the lateral direction, which results in a resonant mode with narrow bandwidth. On the other hand, the bright mode is associated with the weak Fabry-P\'erot mode induced by the index difference in the vertical direction, which results in a mode with nearly flat-band spectrum. The interference of these two modes result in an asymmetric line shape of the spectrum~\cite{Zhou2009, Wu2013}. Moreover, this phenomenon could confine the optical mode in the structure well and result in a high Q factor, which is beneficial for the BIC laser operation with very narrow linewidth (red curve). 

Fig.~\ref{Active}(b) shows the detail characterizations of lasing behaviour. The black curve is the conventional plot of output power as a function of pump energy density, so we can estimate the threshold is around 17 $\mu$J/cm$^2$. The blue curve represents the corresponding optical spectrum linewidth changing, the sharp narrowing of spectrum also indicates a coherence emission is achieved when the laser is operated above threshold. Both of two kinds of characterizations indicate the BIC laser has been successfully demonstrated.   

Fig.~\ref{Active}(c) and (d) are the electric field intensity distributions of resonator in ($x$, $y$) and ($x$, $z$) planes, respectively. When the resonator is operated above threshold, a large mode overlaps with the input pump contributing to a simple and efficient coupling of the input light, as suggested by the high intensity interfere fringes. From the side view (Fig.~\ref{Active}(d)), it is clear how the light is predominantly confined within the grating bars and there is exactly one intensity lobe within the interval of neighboring bars. In the middle, higher intensity stripe indicates the gain medium region, where we can find the vertical confinement of this device is much stronger than the confinement typically found in conventional DBRs-based vertical cavity emitting lasers. Since the double gratings provide the largest power reflectance, all diffraction orders disappear except the 0th order. In addition, the HCGs bring a higher effective refraction index and much higher than the surrounding environment, leading to strong lateral confinement for the lateral modes.  

\section{Sensing function for gas detection}
\begin{figure}
\centering
\includegraphics[width=0.6\linewidth,clip=true]{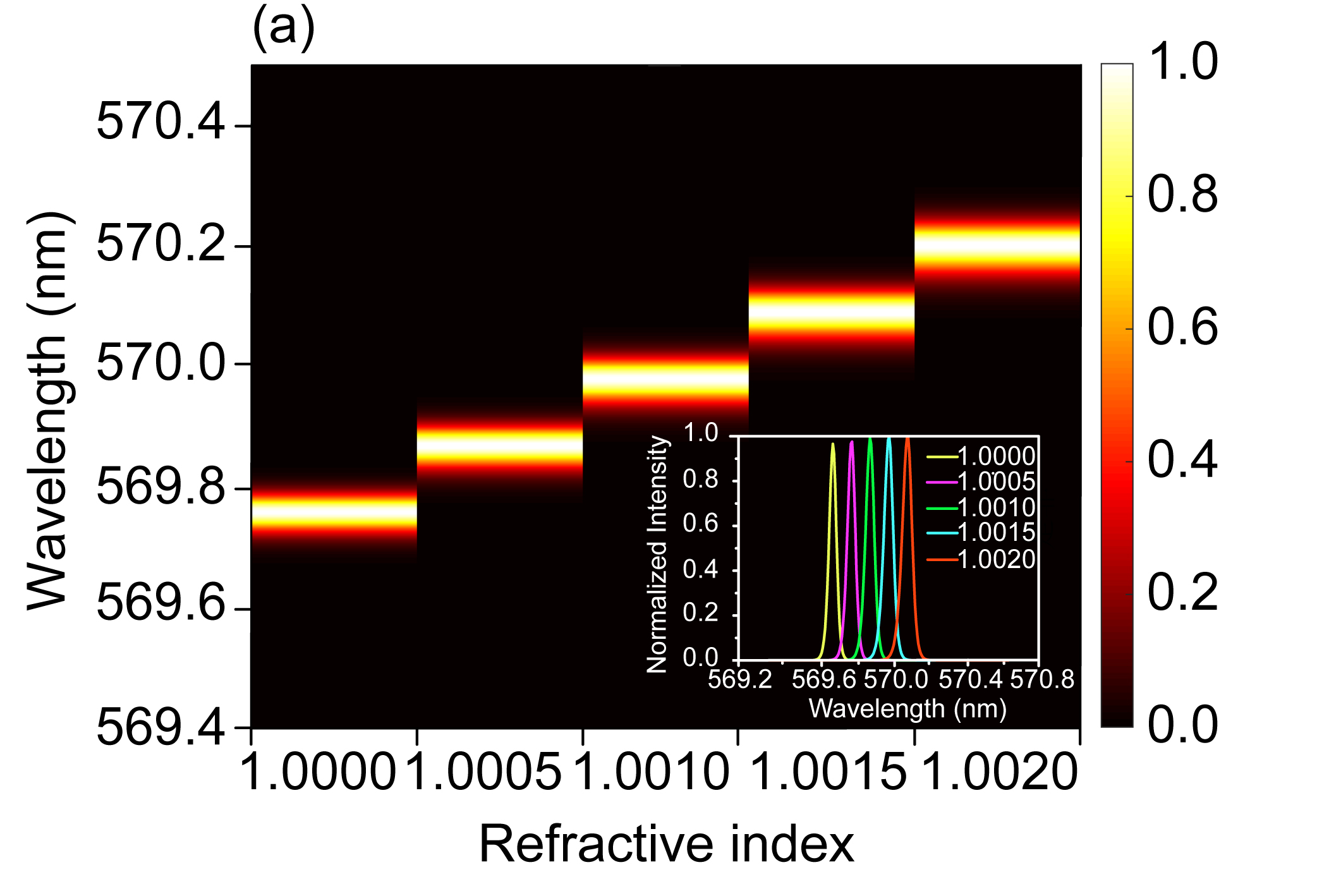}
\includegraphics[width=0.6\linewidth,clip=true]{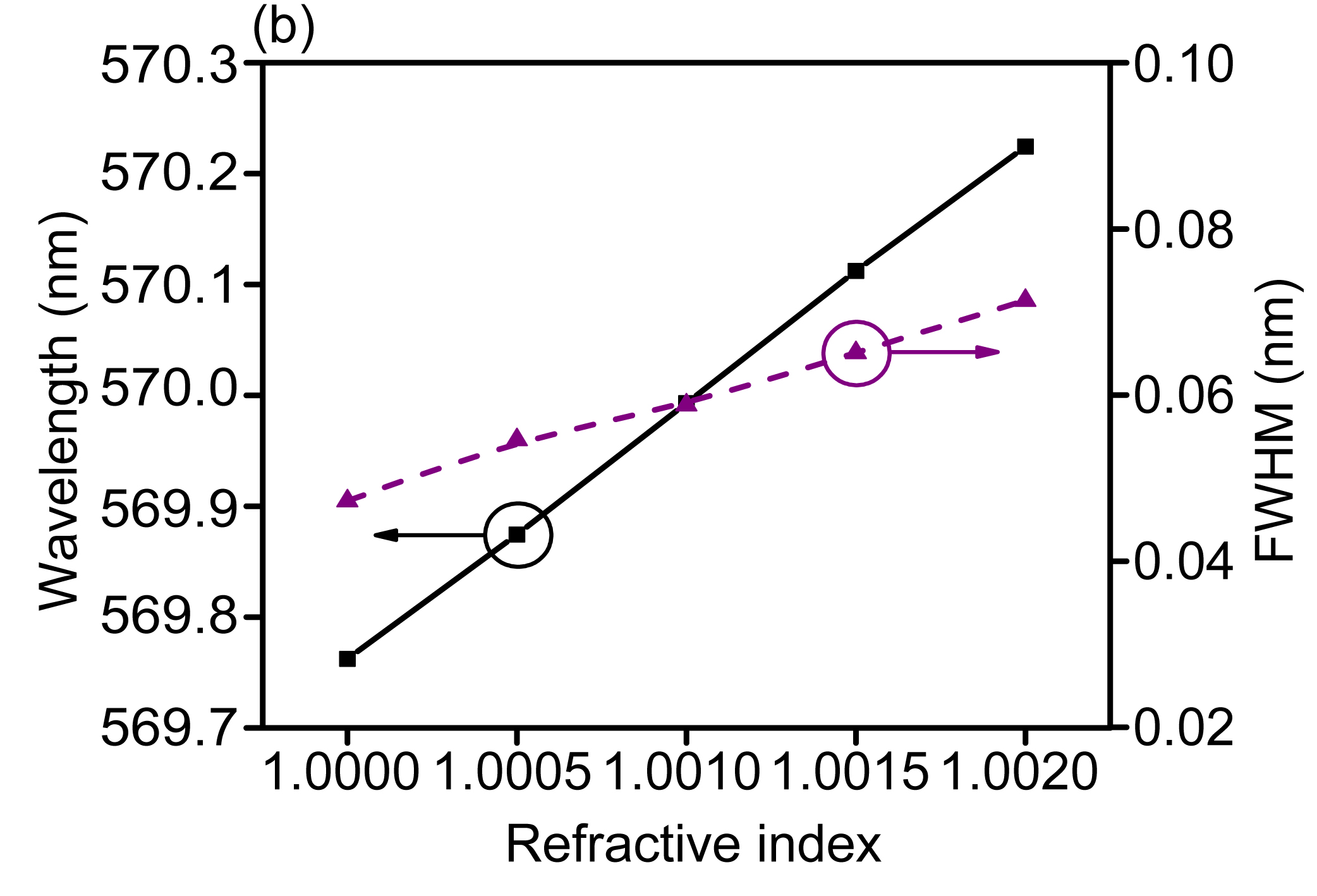}
\caption{(a) Mapping of emission wavelength shift with the variation of environment refractive index, inset: full lasing spectra for different refractive indexes; (b) linear relationship between emission wavelength shift and environment refractive index (black curve), and the corresponding spectrum linewidth changing with refractive index.}
\label{Shift}
\end{figure}

High-$Q$ resonators with narrow resonance line widths and long photon storage time, have been considered as ideal candidates for sensors with enhanced detection sensitivity~\cite{Armani2003, Zhu2010}. Therefore, we carry out quantitative calculations of the quasi-BIC laser intrinsic sensitivity upon changes of the surrounding refractive index. We first reproduced a gas environment and immersed the resonator in a atmosphere of known refractive index value, then performing a refractive index change from 1.0000 to 1.0020. Fig.~\ref{Shift}(a) displays the full lasing spectra of the device \textit{vs.} variation of refractive index, where it is easily observe a shift of the emission wavelength towards a longer wavelengths, even though the refractive index increase step was just of the order of 5 $\times$ 10$^{-4}$. The inset exhibits the laser spectra upon change of the gas environment, where we notice a single wavelength shifting without any other obvious changing. This also can be found from Fig.~\ref{Shift}(b), which presents the linear dependence of emission peaks on the refractive index (black curve), and the linewidth variation (dash curve). It is further confirmed that the change of refractive index results in only the shift of lasing wavelength without affecting the lasing state. This simple linear relationship between lasing wavelength shift and the variation of environment confirms the sensing capability of the proposed microlaser structure. 

Specifically, the sensing performance of the device was evaluated by estimating the sensitivity ($S_{\lambda}$) and figure of merit ($FOM_{\lambda}$), which can be calculated through using the following formulas:

\begin{eqnarray}
S_{\lambda} = \frac{\Delta \lambda}{\Delta n}\, , \\
FOM_{\lambda} = \frac{S_\lambda}{\gamma}\,
\end{eqnarray}

\noindent where $\Delta \lambda$ represents the emission wavelength shifting, and $\Delta$\textit{n} is the change of refractive index. $\gamma$ is the emission spectrum linewidth measured at its full width at half maximum. FOM parameter is another important and more comprehensive benchmark to evaluate sensing performance~\cite{Zhou2019, Huang2016}. From the results in Fig.~\ref{Shift}(b), we calculated the $S_{\lambda}$ and $FOM_{\lambda}$, which are 221 nm/RIU and 4420 RIU$^{-1}$, respectively. Very importantly, the achieved FOM value resulted much higher than the theoretical result for localized surface plasmon resonance (LSPR) sensors ($\sim$ 20)~\cite{Bingham2010}, photonic crystal (PhC) cavity~\cite{Liu2017}, microring~\cite{Yen2006} and fiber Bragg grating (FBG) sensors~\cite{Li2019}. It is also comparable with that of state-of-the-art symmetry guided-mode resonance based sensors~\cite{Zhou2019}. In addition, the organic gain enabled laser design leads to a coherent light source, which can either be detected at optical far field, or be easily coupled into fiber system. Thus, this flexible operation will facilitate the experimental measurements. In addition, the simple structure design will effectively reduce the cost of fabrication, making the sensor more appealing.
 
\section{Conclusion}
In conclusion, we have presented the design and numerical characterization of a compact quasi-BIC-laser based sensor for detecting of gas environment changing. The sensor is designed by using double Si$_3$N$_4$ HCG grating layers and organic R6G dye molecule doped SiO$_2$ layer located between the grating layers. The proposed unique hybrid design enables stimulated vertical emission from the organic gain medium embedded in the HCG stripes. Optimizations concerning HCGs and microcavity configuration designed for a 570 nm resonant wavelength have been performed. Our design features a high-Q resonance, which enables a narrow-linewidth laser emission of 0.05 nm. Therefore, a pronounced BIC lasing behaviour with narrow spectrum linewidth have been realized. The emitted light is close to the plane wave in air, which indicates a single lobe far field pattern for laser beam distribution. The sensing mechanism can be reasonably explained through the guided-mode resonance effect originated by constructive interference between the guided modes of the slab grating and the strong reflection modes. Finally, we also investigated the bulk sensing performance by inserting the device in some specific environment with different refractive index values, demonstrating remarkable sensitivity and figure of merit estimated equal to 221 nm/RIU and 4420 RIU$^{-1}$, respectively. Especially the figure of merit value outperforms recently reported GRM devices~\cite{Wan2017, Zhou2019} and other state-of-the-art passive sensors~\cite{Liu2017}. In view of these achievements, it is believed that this new platform could improve the current technology in remote gas sensing, e. g. warning from dangerous chemical gas.

\section*{Acknowledgment}
We are grateful to Prof. Yang Li from Tsinghua University for useful discussions. T. Wang acknowledges financial support from the National Natural Science Foundation of China (Grant No. 61804036) and Zhejiang Province Commonweal Project (Grant No. LGJ20A040001). I. De Leon acknowledges the support of the Federico Baur Endowed Chair in Nanotechnology. L. Peng thank the support from the Natural Science Foundation of China (Grant No. 61875051) and Zhejiang Province Natural Science Foundation (LR21F010002). F. Gao thank the support from the Natural Science Foundation of China (Grant No. 61801426) and Zhejiang Provincial Natural Science Foundation (Grant No. Z20F010018). X. Lin thank the support from the Fundamental Research Funds for the Central Universities and Zhejiang University Global Partnership Fund.



\end{document}